\documentclass[aps,prb,twocolumn, reprint,groupedaddress,amsmath,amssymb]{revtex4-1}

\usepackage{graphicx}
\usepackage{bm}

\usepackage{natbib} 

\usepackage{color}

\usepackage{float}
\begin{document}


\title{Two orders of magnitude reduction in silicon membrane thermal conductivity by resonance hybridizations}


\author{Hossein Honarvar}
 
\author{Mahmoud I. Hussein}%
 \email{mih@colorado.edu}

\affiliation{Department of Aerospace Engineering Sciences, University of Colorado Boulder, Boulder, Colorado 80309, USA}


\date{\today}

\begin{abstract}
The thermal conductivity of a freestanding single-crystal silicon membrane may be reduced significantly by attaching nanoscale pillars on one or both surfaces.~Atomic resonances of the nanopillars locally and intrinsically couple with the base membrane phonon modes causing these modes to hybridize and flatten at each coupling location in the phonon band structure.~The ensuing group velocity reductions, which in principle may be tuned to take place across silicon's full spectrum, lead to a lowering of the in-plane thermal conductivity in the base membrane.~Using equilibrium molecular dynamics simulations, we report a staggering two orders of magnitude reduction in the thermal conductivity at room temperature by this mechanism.     
\end{abstract}


\maketitle

The emerging field of~\it{phononics} \rm seeks to elucidate the nature of phonon dynamics in both conventional and artificially structured materials and use this knowledge to extend the boundaries of physical response at either the material or structural/device level or both~[\onlinecite{Hussein_AIPA_2011,*Hussein_AIPA_2014}].~This process targets acoustic, elastic, and/or thermal properties and usually involves the investigation and utilization of complex wave mechanisms encompassing one or more of a diverse range of phenomena such as dispersion, resonances, dissipation, and nonlinear interactions~[\onlinecite{Hussein_JVA_2014}]. \\     
\indent In the subfield of~\it{nanophononics}\rm, an intensely active area of research is the search for strategies for reducing a material's thermal conductivity~[\onlinecite{Chen_2000,*Balandin_2005,*Li_RMP_2012,*cahill2014nanoscale_short,*volz2016_short}], and in particular strategies that would not deteriorate the electrical properties~[\onlinecite{Hopkins_2011,*Alaie_2015}].\footnote{The advent of nanotechnology~[\onlinecite{Wolf_2015}] and nanofabrication techniques~[\onlinecite{Biswas_2012}] is providing a powerful enabling tool in this search.}~Attaining a low value of the thermal conductivity $k$ and simultaneously high values of the electrical conductivity $\sigma$ and the Seebeck coefficient $S$ is strongly desired in thermoelectric materials$-$materials that convert heat in the form of a temperature difference to electric energy, or, conversely, use electricity to provide heating or cooling~[\onlinecite{chen2003recent,*dresselhaus2007new,*VineisAM2010}].~The performance of a thermoelectric material is measured by a figure of merit defined as $ZT = \sigma S^{2} T/k$, where $T$ is the absolute temperature and $k$ is the sum of a lattice component $k_l$ and an electrical component $k_e$~[\onlinecite{Rowe2005}].\\
\indent Over the past few decades, a large research effort has focused on semiconductors where $k_l\gg k_e$. The prevailing approach to increasing $ZT$ in this class of materials is to introduce small and closely spaced features (such as holes, particles, and/or interfaces) within the internal domain of the material to scatter the heat-carrying phonons and consequently reduce the lattice thermal conductivity, e.g., see Ref.~[\onlinecite{Liu2012nanocomp,*Kanatzidis2012}].~This strategy, however, faces the challenge that the scatterers are also likely to impede the transfer of electrons and thus negate any possibility of substantial increase in $ZT$.~Another approach is the use of superlattices~[\onlinecite{Cleland_2001,*McGaughey_2006,*Landry_2008,*Luckyanova_2012_short}] or nanophononic crystals~[\onlinecite{Hopkins_2011,*Alaie_2015},\onlinecite{Gillet_2009,*Tang_2010,*He_2011,*Davis_2011,*Robillard_2011,*Yang_2014},\onlinecite{Yu_2010}] where the aim is to use Bragg scattering to open up phonon band gaps and reduce the group velocities by flattening the dispersion curves.~A practical disadvantage to this route, however, is that the surfaces of the periodic features, e.g., the layers, holes or inclusions, need to be considerably smooth to preserve the phase information required for the Bragg effects to take place$-$especially when the features are of relatively large sizes compared to the phonon wavelengths.~An even stronger drawback is that the degree and intensity of group-velocity reduction is rather limited and cannot be enhanced beyond what the available Bragg interference patterns can provide.~One other promising avenue is through dimensionality reduction, e.g., considering thermal transport along a nanowire~[\onlinecite{boukai2008silicon}].~This introduces phonon confinement and strong phonon scattering at the free surfaces, especially when roughened or oxidized~[\onlinecite{Neogi_2015}].~Yet this approach too, when used alone, is relatively limited in its capacity to lowering the thermal conductivity without excessive reduction in the size of the smallest dimension.\\
\begin{figure}[b!]
	\includegraphics{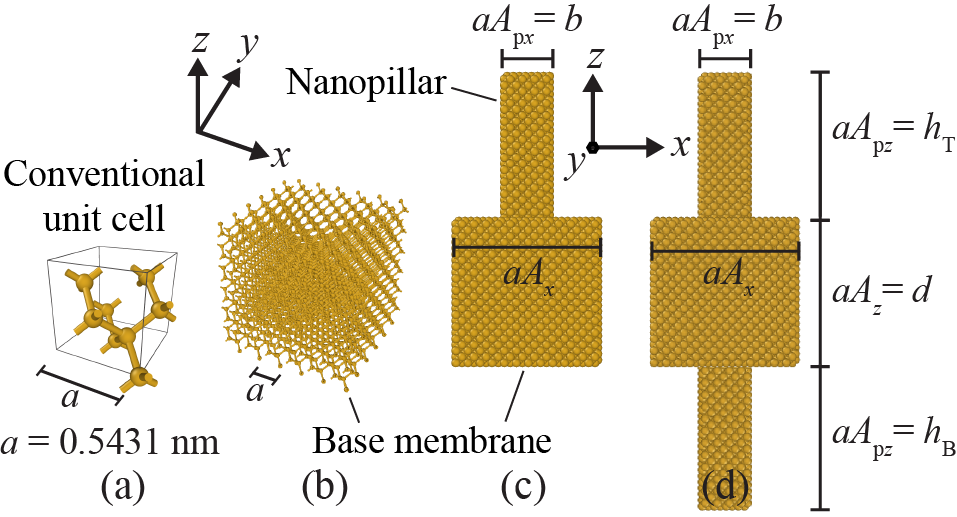} 
	\caption{\label{fig:fig_1} (a) Conventional 8-atom unit cell for silicon and unit cells for a (b) uniform membrane, (c) single-pillared NPM, and (d) double-pillared NPM.}
\end{figure}
\begin{figure*}[t!]
	\includegraphics{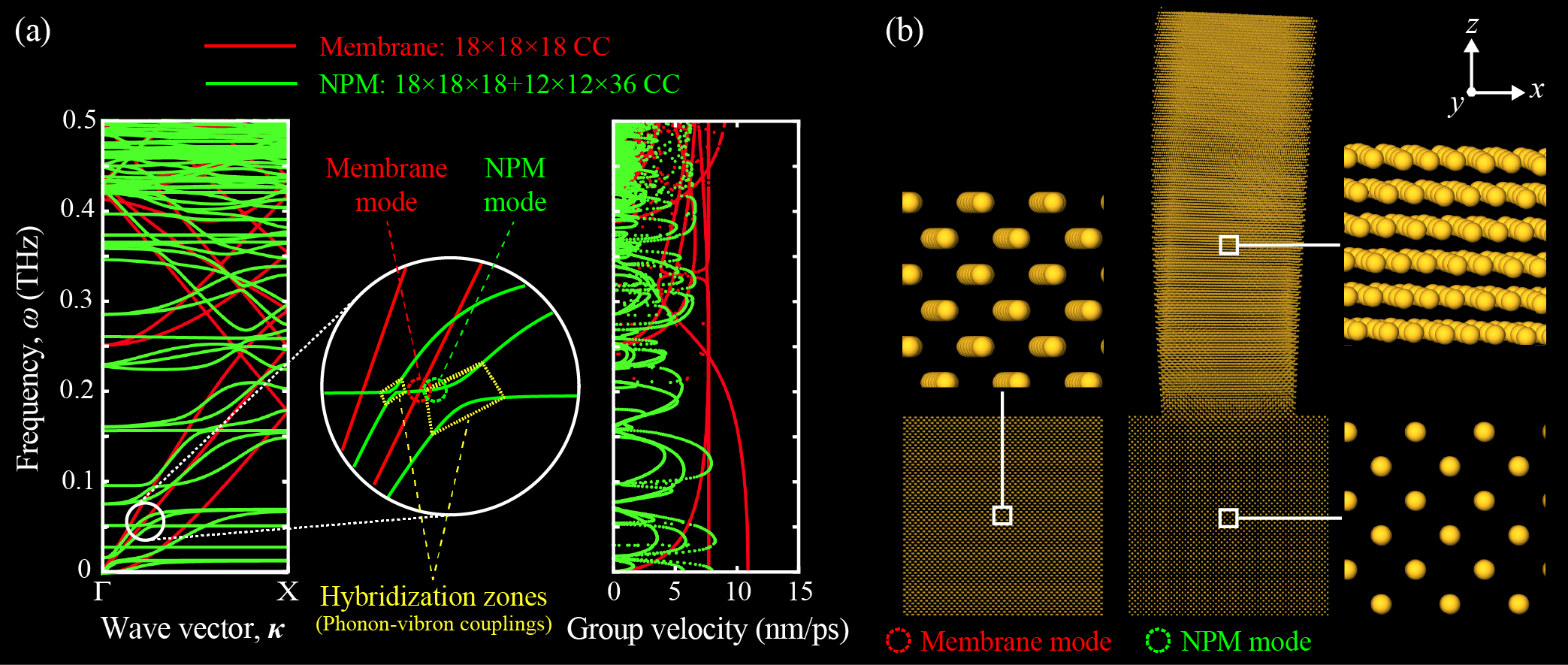} 
	\caption{\label{fig:fig_2} Illustration of the resonance hybridization phenomenon from a lattice dynamics perspective.~(a) Phonon band structure and group velocity distribution of a silicon membrane with (green) or without (red) silicon nanopillars standing on one surface.~(b) Uniform membrane atomic displacements for a heat carrying phonon mode in the acoustic regime contrasted to NPM atomic displacements of the same mode upon resonance hybridization.~Significant motion within the uniform membrane is seen.~In contrast, the atomic displacements of the NPM hybridized mode reveals localized nanopillar motion and almost `thermal silence' in the base membrane portion.~In (a), a zoom-in is provided for two hybridization zones including the one illustrated in (b).~A magnification factor of 2000 is applied to the atomic displacements in the mode-shape images.}
\end{figure*}
\indent In our group at CU-Boulder, we have been investigating a fundamentally different paradigm for increasing $ZT$.~Instead of depending on boundary-type scattering (internal or external), Bragg interferences, and/or phonon confinement as leading mechanisms for lowering $k_l$, we employ local resonances~[\onlinecite{PhysRevLett.112.055505,honarvar2016PRB,honarvar2016APL}].~In this concept, termed \it{nanophononic metamaterial} \rm (NPM), nanoscale resonating substructures are intrinsically introduced to a conventional semiconducting material which acts as the prime thermoelectric medium.~The purpose of these substructures is not to generate subwavelength band gaps or create negative long-wave effective properties as is the case for locally resonant electromagnetic~[\onlinecite{Pendry_IEEE_1999,*Smith_PRL_2000}], acoustic~[\onlinecite{liu2000locally}] and elastic~[\onlinecite{liu2002elasticMM,*pennec2008low,*wu2008evidence}] metamaterials, but to reduce the phonon group velocities of the underlying semiconducting material in order to significantly reduce its thermal conductivity.~The substructure resonances may be designed to couple with all or most of the heat-carrying phonon modes across the full spectrum of the host medium.~\footnote{Another distinction for NPMs compared to locally resonant electromagnetic, acoustic and elastic metamaterials is that they draw on their unique properties for the function intended across the full spectrum and not just within the subwavelength regime.}~This atomic-scale coupling mechanism gives rise to a \it{resonance hybridization} \rm between pairs of the wavenumber-independent vibration modes of the local substructure (vibrons) and wavenumber-dependent wave modes of the host medium (phonons)$-$the stronger the couplings, the sharper the curve flattenings and the stronger the reductions in the group velocities.~In the limit, the number of hybridizing vibrons is three times the number of atoms in a unit nanoresonator. \\
\begin{figure*}[ht]
	\includegraphics{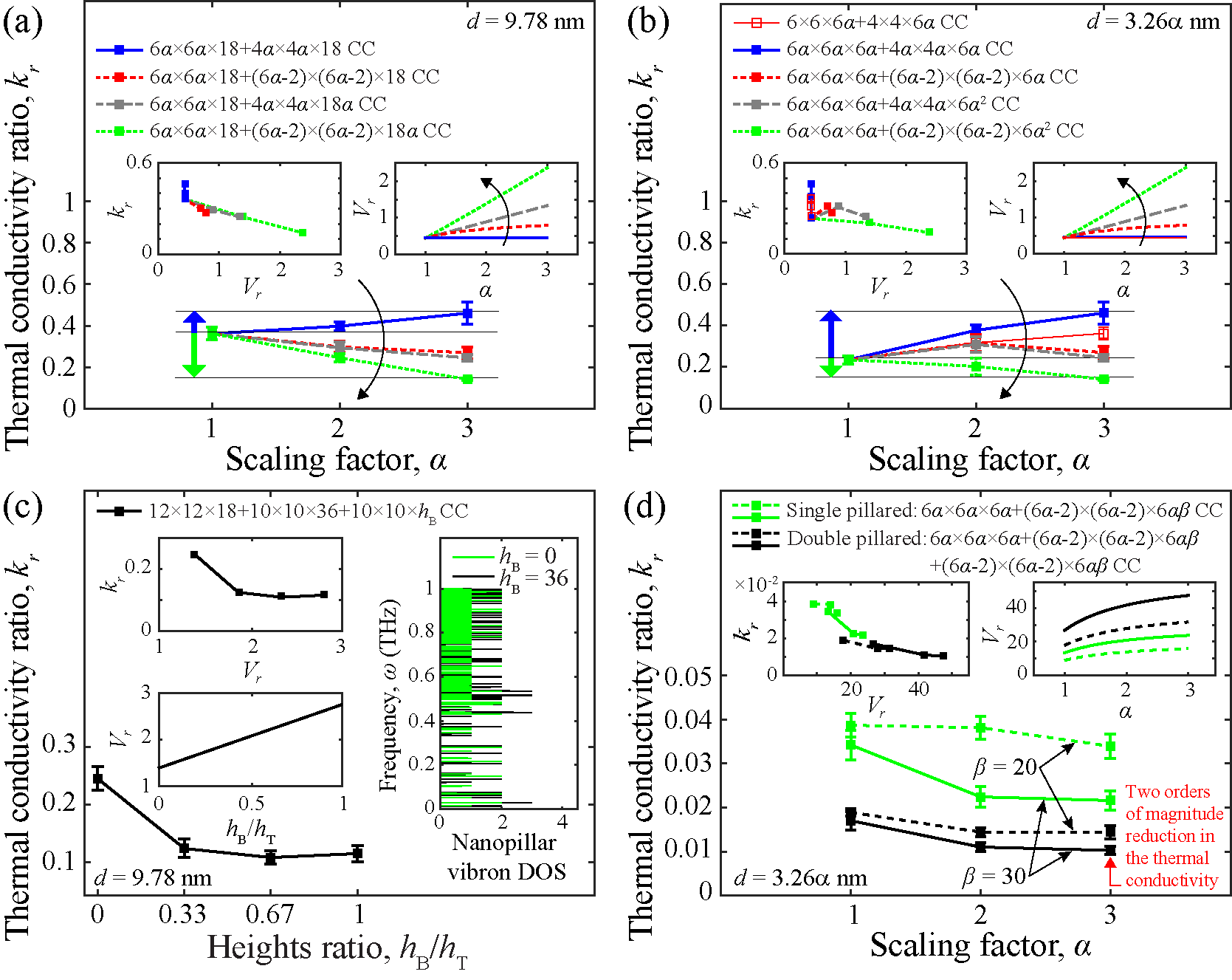} 
	\caption{\label{fig:fig_3} Thermal conductivity ratio $k_{r}$ versus the scaling factor $\alpha$ in (a), (b) and (d) and versus the ratio of top and bottom nanopillar heights $h_{\textrm{T}}/h_{\textrm{B}}$ in (c).~A range of unit-cell configurations with different $\alpha$-dependencies is considered.~A membrane thickness of 6, 12, and 18 CC correspond to 3.26, 6.52, and 9.78 nm, respectively.~Thick arrows in (a) and (b) represent maximum upward and downward changes among the cases considered.~In each of (a), (b) and (d), the left inset shows $k_{r}$ versus the nanopillar-to-base membrane volumetric ratio $V_r$ (error bars not shown for brevity), and the right inset plots $V_r$ versus $\alpha$.~In (c), the top left inset shows $k_r$ versus $V_r$ as $h_{\textrm{T}}/h_{\textrm{B}}$ increases, the bottom left inset plots $V_r$ as a function of $h_{\textrm{T}}/h_{\textrm{B}}$, and the right inset shows the vibrons DOS of a single- versus double-pillared NPM unit cell as directly extracted from the band structure.~All trends clearly show that the performance of an NPM in reducing $k_r$ is directly dependent on $V_r$ and the type of $\alpha$-dependency.~In the double-pillared configuration considered in (d), a two orders of magnitude reduction in the thermal conductivity is recorded.}
\end{figure*}
\indent A candidate configuration of an NPM consists of an array, or a forest, of silicon nanopillars distributed on the surface(s) of a freestanding silicon membrane with no interior scatterers~[\onlinecite{PhysRevLett.112.055505,honarvar2016PRB,honarvar2016APL}].~Here the nanopillars act as the resonating substructures.~Since the nanopillars are located external to the main body of the membrane, the scattering of electrons is minimized.~Compared to all conventional phonon scattering-based approaches, this new route therefore provides the unique advantage of practically decoupling the lattice thermal conductivity from the electrical conductivity and the Seebeck coefficient$-$which is essential to creating significant improvements in $ZT$.~And compared to superlattices and nanophononic crystals, an NPM in general has two advantages:~(i) the structural features do not need to be periodic or smooth (because the resonance hybridization phenomenon is independent of periodicity and robust to perturbations in phase), and (ii) the degree and intensity of group-velocity reductions may be continuously enhanced by simply increasing the size of the nanoresonators~[\onlinecite{honarvar2016APL}].\footnote{A monotonic improvement in performance will take effect with increasing nanoresonator size until the characteristic length scales of the nanoresonator, or the unit cell as a whole, exceed the full span of the phonon mean free path distribution$-$beyond this point, scattering mechanisms will dominate and the probability of occurrences of resonance hybridizations will decay as a result.}~Finally, a nanopillared freestanding membrane naturally exhibits dimensionality reduction (compared to the bulk form).~Therefore, the powerful rewards of resonance hybridizations are gained over and above the benefits of phonon confinement and/or surface roughness.~In light of these impressive characteristics that are unprecedented in thermal transport, the NPM concept in the form of a nanopillared membrane is poised to enable thermoelectric energy conversion at record high performance, while using a low-cost and practical base material like silicon.\\
\indent In a recent study involving nanopillars on one surface, it was shown that the performance of this membrane-based NPM configuration is highly dependent on (i) the relative volumetric size of the nanopillar with respect to the base membrane within the unit cell (this quantity is denoted $V_r$) and (ii) the overall size of the unit cell (including both the base membrane and nanopillar portions)~[\onlinecite{honarvar2016APL}].~While the first dependency provides a controllable design parameter (which is an advantage as mentioned above), the second was shown to pose a challenge because unless $V_r$ is relatively high to start with, the extent of the thermal conductivity reduction may deteriorate as the overall unit-cell size is proportionally scaled up~[\onlinecite{honarvar2016APL}].~In this Letter, we explore the possibility of ``compensating" this loss in performance by increasing the nanopillar size at a higher rate than the base membrane as we progressively examine larger unit cells.~By following this path, we demonstrate that it is possible to even reverse the trend and achieve improved performance with upscaling.\\ 
\indent We investigate two prime freestanding NPM configurations:~a membrane with nanopillars (i) on one surface, and (ii) on each of the surfaces.~In all cases, both base membrane and nanopillar(s) are made of defect-free single-crystal silicon.~Fig.~\ref{fig:fig_1} displays the unit cells of these two configurations as well as the structure of a conventional cell (CC) and a unit cell of a corresponding uniform (unpillared) membrane.~The geometry of a membrane with nanopillars on each surface is represented as $aA_{x}\times aA_{y}\times d+b\times b\times h_{\rm T}+b\times b\times h_{\rm B}$ which may be converted to CC by dividing each dimension by $a$. Each of the last two terms in this representation is dropped as needed when representing an unpillared surface.~All geometric parameters are pictorially defined in Fig.~\ref{fig:fig_1}.\\
\indent Equilibrium molecular dynamics (EMD) simulations and the Green-Kubo method~[\onlinecite{Zwanzig_1965,*Ladd_1986,*volz2000molecular,*Schelling_2002}] are used for the majority of the thermal conductivity calculations; see~[\onlinecite{honarvar2016PRB}] and~[\onlinecite{honarvar2016APL}] for implementation details.~Lattice dynamics calculations and the Boltzmann transport equation following the single-mode relaxation time approximation are also utilized and this prediction method is described in~[\onlinecite{PhysRevLett.112.055505}].~In all the analysis, room temperature, $T = 300$ K, is assumed and the Stillinger-Weber empirical potential is used to represent the interatomic interactions~[\onlinecite{stillinger1985computer}]. 
\indent In Fig.~\ref{fig:fig_2}, we provide a demonstration of the resonance hybridization phenomenon as manifested in the phonon band structure, group-velocity distributions, and associated mode shapes.~For this purpose, we consider an $ 18\times 18\times 18+12\times 12\times 36$ CC NPM unit cell (consisting of 88128 atoms) and use the reduced Bloch mode expansion (RBME) technique~[\onlinecite{Hussein_PRSA_2009}] to solve the problem over the 0-0.5 THz range.~The effects of the vibron-phonon mode coupling phenomenon are clearly displayed. \\
\begin{figure}
	\includegraphics{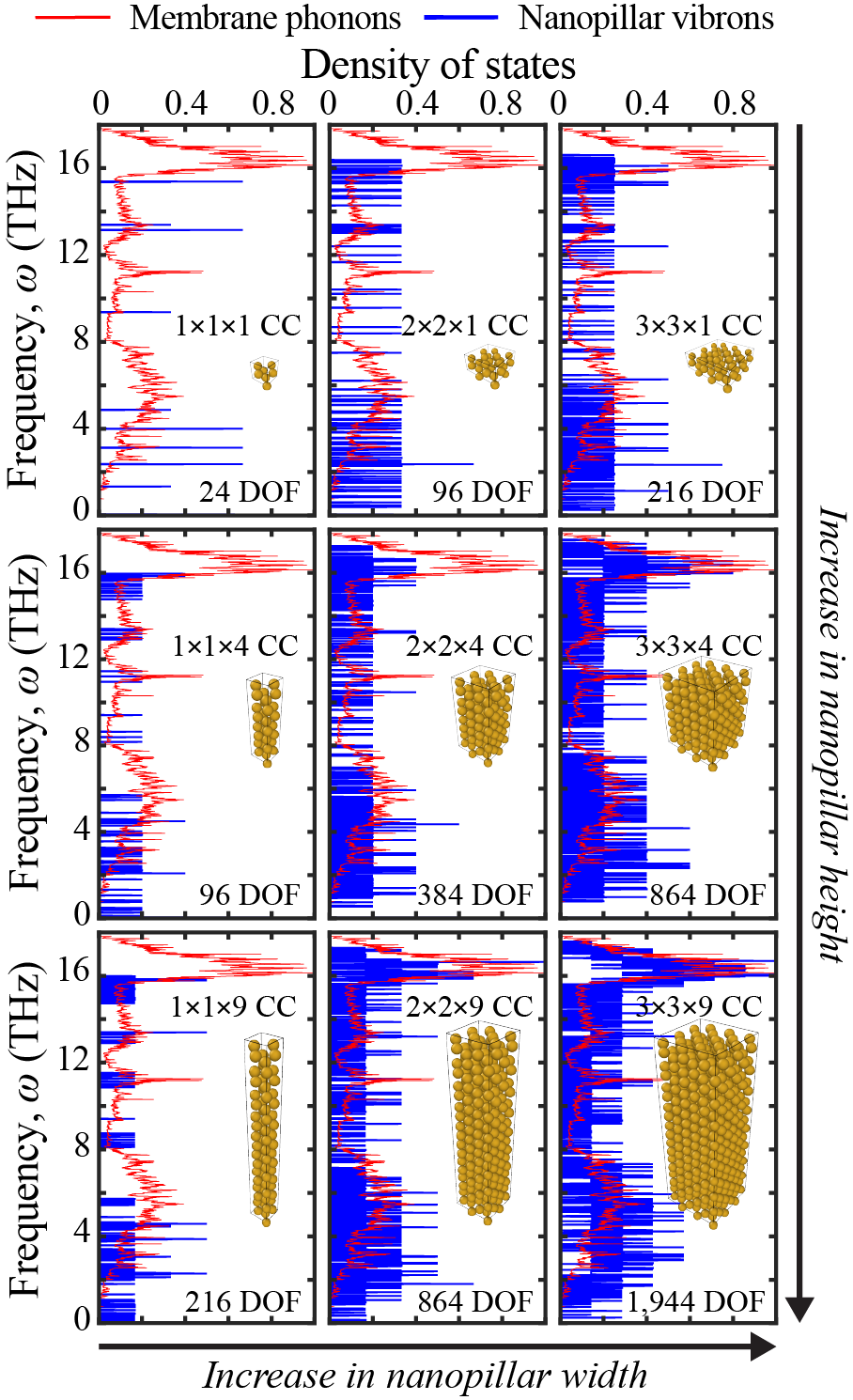} 
	\caption{\label{fig:fig_4} Direct correlation between silicon membrane phonons DOS (considering a $ 6\times 6\times 6$ CC membrane) and silicon nanopillar vibrons DOS for various nanopillar side lengths and heights.~Unlike in~Fig.~\ref{fig:fig_3}c, here the latter quantities are obtained by considering the nanopillar as an independent nanostructure with free boundary conditions.~All quantities are normalized with respect to their maximum values.}
\end{figure}
\indent In Fig.~\ref{fig:fig_3}, we investigate using EMD simulations the reduction in the thermal conductivity for various NPM geometries and sizes.~This reduction is represented by $k_r$, the thermal conductivity of a membrane with nanopillar(s) divided by that of a corresponding uniform membrane.~Key factors in this analysis are (i) $V_r$ and (ii) the manner by which the geometry of the different NPM components grows with size relative to each other.~Using $\alpha$ as a size parameter, we refer to the latter factor as the ``$\alpha$-dependency".~A graphical representation of the various $\alpha$-dependencies considered is provided in the insets of the figure in the form of $V_r$ vs. $\alpha$, $V_r$ vs. $h_{\rm T}/h_{\rm B}$, and $k_r$ vs. $V_r$. \\
\indent The results in Figs.~\ref{fig:fig_3}a and ~\ref{fig:fig_3}b show that regardless of the $\alpha$-dependency, $k_r$ increases with size when $V_r$ is constant.~We have shown earlier that this behavior is driven by the manner by which the resonance hybridizations affect the group velocities as a unit cell is proportionally upscaled in size~[\onlinecite{honarvar2016APL}].~For $k_r$ to maintain its value with $\alpha$, or possibly even drop in value, we need to introduce a \it compensatory effect \rm in the $\alpha$-dependency; that is, to select the dependency in a manner such that $V_r$ increases as $\alpha$ increases.~The strength of this effect is measured by $\gamma=dV_r/d\alpha$.~We observe that indeed $k_r$ drops, as desired, for the cases exhibiting a compensatory $\alpha$-dependency, e.g., $k_r$ drops from 0.36 at $\alpha=1$ to 0.14 at $\alpha=3$ for the $ 6\alpha\times 6\alpha\times 18+(6\alpha-2)\times (6\alpha-2)\times 18\alpha$ CC model for which $V_r=\alpha(1-1/3\alpha)^2$ and $\gamma=(1-1/9\alpha^2)$.\\
\indent In Fig.~\ref{fig:fig_3}c, we examine the effect of adding a second nanopillar (at the bottom) of the membrane.~The right inset demonstrates that this significantly increases the spread in the vibron density of states (DOS) spectrum, and as a result $k_r$ drops from 0.25 to 0.12.~Pushing the compensatory effect further, we show in Fig.~\ref{fig:fig_3}d a reduction in the thermal conductivity by a factor of 45 for a single nanopillar and a factor of 97 for double nanopillars for a membrane 9.78-nm thick with each nanopillar extending up to 293.3 nm in height.~The extent of these reductions is unprecedented in the literature, and yet more reductions are possible with further increases in nanopillar size and potentially other treatments such as optimized alloying.\\
\indent Finally, we show in Fig.~\ref{fig:fig_4} the effects of the size and geometry of a nanopillar on the distribution of the vibrons DOS and how it correlates with that of the phonons DOS of the underlying base membrane.~It is observed that for the same number of atoms, a wider nanopillar provides a more spread-out local resonances spectrum than a tall nanopillar.~The higher the vibron densities and the more conforming to the phonons distribution, the more effective the resonance hybridization phenomenon$-$especially at low frequencies down to the limit of existing wavelengths.~For the largest nanopillar, we clearly see a perfect conformity between the two distributions owing to the fact that both membrane and nanopillar are made of the same material. \\ 
\indent From our calculations, the thermal conductivity of a 9.78-nm thick uniform silicon membrane is 3.7 times lower than the bulk form.~Multiplying this by the factor of 97 for the best double-nanopillar case reported above gives a total factor of nearly 360.~By linear extrapolation of the observed $k_r$ versus $\alpha$ trend for this high performing configuration, a 19.55-nm thick NPM would exhibit a thermal conductivity reduction by a factor of 119 with respect to a corresponding uniform membrane and a factor of 290 with respect to bulk silicon.~Assuming a one-to-one mapping between the thermal conductivity reduction and the increase in $ZT$$-$as demonstrated by the experimental characterization of similarly sized freestanding silicon membranes~[\onlinecite{Yu_2010}] and silicon nanowires~[\onlinecite{boukai2008silicon}], and given that the $ZT$ of bulk silicon at $T=300$ K is 0.01~[\onlinecite{weber_1991}], we obtain a projected room-temperature $ZT$ value of 2.9.~This is significantly higher than any previously reported $ZT$ value at any temperature, not only for silicon but for thermoelectric materials in general.\\  
\indent This research was partially supported by the National Science Foundation (NSF) CAREER Grant No.~1254931 and the Smead Faculty Fellowship program.~The authors thank Mr. D. Krattiger and Dr.~J. Gale (Curtin University) for their helpful tips concerning LD calculations.~This work utilized the Janus supercomputer, which is supported by NSF Grant No. CNS-0821794 and the University of Colorado Boulder. 

\bibliography{Ref_NPM_Design}

\end{document}